\documentclass[prb,twocolumn]{revtex4-1}

\usepackage{amsmath, amsthm, amssymb}
\usepackage{amsfonts}
\usepackage{graphicx}
\usepackage{gensymb}

\newcommand{\bq}{\begin{equation}}
\newcommand{\ee}{\end{equation}}
\newcommand{\C}{\text{C}}

\begin{document}

\title{Reversible temperature exchange upon thermal contact}

\author{Eugene G. Mishchenko}
\affiliation{Department of Physics and Astronomy, University of Utah, Salt Lake
City, UT 84112, USA}

\author{Paul F. Pshenichka}
\affiliation{Lyceum \#1, Chernivtsi 58001, Ukraine }

\begin{abstract}
According to a well-known principle of  thermodynamics, the
transfer of heat between two bodies is reversible when their
temperatures are infinitesimally close. As we demonstrate, a
little-known alternative exists: two bodies with temperatures
different by an arbitrary amount can completely exchange their
temperatures in a reversible way if split into infinitesimal
parts that are brought into thermal contact sequentially.
\end{abstract}


\maketitle

\section{Introduction}

This story dates back almost 30 years, when one of us, a
high school teacher, found a curious  note in an obscure Soviet
book about a fascinating phenomenon.\cite{Mak}  He called the
subject to the attention of his students, the other author
being among them. As intriguing as the problem appeared, its
complete solution eluded us for many years. And while we
clearly cannot take credit of inventors of the main principle
here---after all this principle has been used for decades in
commercial heat exchangers\cite{exchanger}---the present
analysis of the problem, to our best understanding, is novel
and in any case not commonly known to physics instructors.

It is hard to come up with a more basic thermodynamic question,
or one that even those who have never studied physics might feel
confident to answer.  Consider equal amounts of icy cold
($0\,\degree\C$) and steaming hot ($100\,\degree\C$) water. One
needs to cool the hot water as much as possible by bringing it
in thermal contact with the cold water, but without actually
mixing them. Heat losses to the environment are neglected.

Simply bringing the two waters into direct thermal contact would
obviously result in the final temperature of $50\,\degree\C$
for both, as long as the specific heat of water is
assumed to be independent of temperature.\cite{capacity} If, however, one first
splits the cold water into two equal amounts and {\it then} brings
them in contact with the entire amount of hot water \textit{one after another}, the result is different.  Indeed,
after the first contact the hot water will cool down to
\begin{equation}
100\,\degree \C\times \frac{2}{3} = 66.67 \,\degree  \C.
\end{equation}
After the second contact, the temperature of the hot water will
be
\begin{equation}
66.67\,\degree \C\times \frac{2}{3} = 44.44\, \degree \C,
\end{equation}
which is well \textit{below} the middle value of $50\,\degree \C$.
Correspondingly, the temperatures of the two parts of the
(initially) cold water will be $66.67\,\degree \C$ and
$44.44\,\degree \C$. Upon the subsequent mixing, they reach the final equilibrium
 temperature of $55.56\,\degree \C$, which together with the
final temperature of the formerly hot water, $44.44\,\degree \C$, adds
up to $100\,\degree \C$.

One does not have to stop at splitting the  cold water into
merely two halves. Suppose that it is  separated into $N$ equal
parts and then, as before, each part is brought into contact
with the whole body of the hot water. After the first contact
the hot water will cool down to
\begin{equation}
\frac{t_0}{1+1/N}, \hspace{0.5cm} t_0 = 100\,\degree \C.
\end{equation}
After the second contact its temperature will go down a bit
more to
\begin{equation}
\frac{t_0}{\left(1+{1}/{N}\right)^2}.
\end{equation}
After all $N$ cold parts have been used, the final temperature
of the hot water will be
\begin{equation}
\frac{t_0}{\left(1+{1}/{N}\right)^N}.
\end{equation}
In the limit of infinitely fine splitting, $N \to \infty$, the
ultimate temperature will be
\begin{equation}
t_0/e = 36.79\,\degree \C.
\end{equation}
Finally,  all parts of the ``cold'' water are mixed together
and, according to the conservation of energy, their final
temperature must be $t_0(1-1/e)=63.21\,\degree \C$,
considerably warmer  than the ``hot'' water. While the
temperature of the latter is the same as the temperature of a
human body, the ``cold'' water is too hot for a human to stand.
But can one do even better?

\begin{figure}[h]
\resizebox{.40\textwidth}{!}{\includegraphics{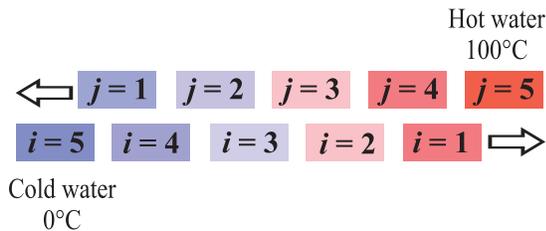}}
\caption{Schematic illustration of the heat exchange process: cold and hot waters are split into $N$
equal parts and brought into contact sequentially with one another, like oncoming trains passing. At each
stop the water parts are allowed full thermal equilibration.
}
\label{Oncoming trains}
\end{figure}

\begin{figure}[h]
\resizebox{.25\textwidth}{!}{\includegraphics{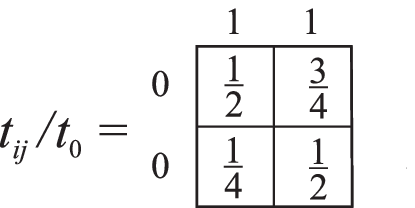}}
\caption{The matrix elements $t_{ij}$ show the temperature of the $i$th part of
the cold water after it makes thermal contact with the $j$th part of the hot water, when both are split into $N=2$ parts.}
\label{Table for two}
\end{figure}

One should not stop at splitting only the cold
water. Suppose that \textit{both} the cold water and the hot water
are split into two equal parts and the first cold part is
brought into thermal contact with both hot parts in sequence;
see Fig.~\ref{Oncoming trains} for the basic principle. The
first cold part is then set aside and the remaining half of the
cold water is brought in contact with the two parts of hot water, now
somewhat cooled down by the passage of the first half of the cold
water. To describe the entire process, it is convenient to
construct the matrix shown in Fig.~\ref{Table for two}.
The element $t_{ij}$ represents the equilibrium temperature
established after the $i$th part of the cold water is brought
in thermal contact with the $j$th part of the hot water. For
example, after the first cold part makes contact with the first
hot part, their common temperature is $t_{11} =(0\,\degree
\C+100\,\degree \C)/2=50\,\degree \C$. After the same cold part
is brought into contact with the second hot part, their
eventual equilibrium temperature is $t_{12} =(50\,\degree
\C+100\,\degree \C)/2=75\,\degree \C$.

When the second cold part moves through the sequence of hot
parts, the initial temperatures of those hot parts are equal to
the equilibrium temperatures achieved after their previous heat
exchanges, given by the elements of the previous row. One
can, therefore, determine the general rule for the construction
of the temperature matrix:
\bq \label{recurrent}
t_{ij}=\frac{1}{2}\Bigl(t_{i-1,j}+t_{i,j-1} \Bigr). \ee

We can extend this recurrence relation to include even the
first row and the first column by introducing  the ``zero'' row
and ``zero'' column,
\begin{equation}
\label{zero row and
column} t_{0j}=t_0, \hspace{0.5cm}t_{i0}=0\,\degree \C,
\end{equation}
which represent the initial temperatures of the hot and cold parts,
respectively.

The final temperature of the (initially) cold water, after all $N^2$
thermal exchanges are completed, is given by the average of the entries in
the last column; and the final temperature of the (initially) hot water is given by
the average of the entries in the last row:
\bq
\label{final temperature}
 t_{\rm cold} (N) =\frac{1}{N} \sum_{i=1}^{N} t_{iN},\hspace{0.5cm}  t_{\rm hot} (N)=\frac{1}{N} \sum_{j=1}^{N}
t_{Nj}.
\ee

Taking the average of  the last row in Fig.~\ref{Table for
two}, one finds the final temperature of the hot water to be,
$t_{\rm hot} (2)=(75\,\degree \C+50\,\degree
\C)/2=37.5\,\degree \C$. Thus, by splitting both waters into just
two parts we have done almost as well as by splitting only one
into infinitely many parts.

\begin{figure}[h]
\resizebox{.25\textwidth}{!}{\includegraphics{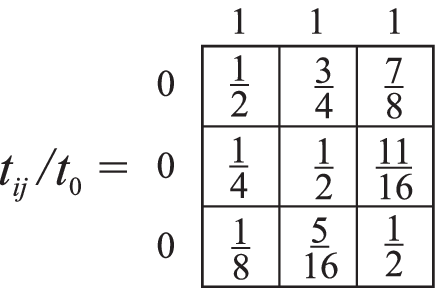}}
\caption{The elements of the temperature matrix for $N=3$ parts.}
\label{Table for three}
\end{figure}

We can do significantly better by making $N=3$ splits. In that
case the matrix of temperature elements is shown in
Fig.~\ref{Table for three} and the final temperature is $t_{\rm
hot} (3)=t_0(1/8+5/16+ 1/2)/3=31.25\,\degree \C$.

The natural question is this:  what is the lowest temperature
that the hot water can be cooled down to if the number of
splittings is increased indefinitely, $N \to \infty$? In
Sec.~II we demonstrate that in this limit a complete temperature inversion
occurs, and that with a very good accuracy the
ultimate temperature obeys the simple formula
\begin{equation}
\label{large n limit}
t_{\rm hot} (N)\approx \frac{t_0}{\sqrt{\pi N}}.
\end{equation}

This result raises the next question: how is it possible that
the outcome of clearly \textit{irreversible} heat exchanges
between bodies with different temperatures is nonetheless a
\textit{reversible} process? Indeed, the simplest definition of a
reversible process in thermodynamics is that the process changes
direction upon an infinitesimal change of the
conditions.\cite{LL}  For example, the heat exchange between two
bodies with temperatures that differ by an infinitesimal amount
is reversible. More rigorously, a process is reversible if
the net change of the entropy of all bodies involved is
zero,\cite{LL,Sam,Rec,Bat} $\Delta S =0$. Correspondingly, when
the bodies brought in thermal contact have a non-infinitesimal
temperature difference, the net entropy change is nonzero and
thus the process is irreversible.

The thermal exchange  with an infinite number of parts ($N\to\infty$)
described in this paper represents a distinct type of reversible process.
Note that a
reversible process facilitated by an infinite number of thermal
baths is well studied in the existing
literature.\cite{Gal,Gup,McL,Hei,Tho,Mir,Cra,Ana,Ana1} We
emphasize, however, that the scenario described here is quite
different: the reversibility does not rely on the presence
of any auxiliary bodies as there are no thermal baths involved.

In the following section we calculate the final temperatures of the hot and cold
waters for arbitrary $N$.  In Sec.~III we treat the continuum
limit, $N\to\infty$, directly.  We then address the
question of the net entropy change in Sec.~IV.

\section{The case of arbitrary $N$}

The recurrence relations (\ref{recurrent}) together with the
``boundary conditions'' (\ref{zero row and column}) determine
the whole matrix $t_{ij}$ for an arbitrary  number $N$ of rows
and columns. Some general properties of this matrix become
clear from the simpler cases of $N=2$ and~3 considered above. We
observe that the diagonal elements all equal $t_0/2$
and that any element and its transpose add up to $t_0$:
\begin{equation}
\label{properties of t}
t_{ij}+t_{ji} = t_0,\hspace{0.5cm} t_{ii}=t_0/2.
\end{equation}
The same properties (\ref{properties of t}) remain valid for
any size of the matrix, as illustrated by Fig.~\ref{Table for
many} for $N=6$.
\begin{figure}[h]
\resizebox{.40\textwidth}{!}{\includegraphics{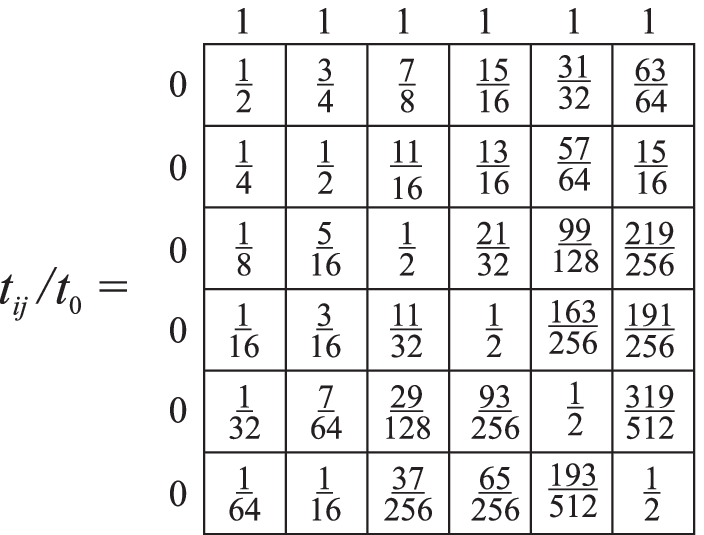}}
\caption{The elements of the temperature matrix $t_{ij}$ for $N=6$ parts.}
\label{Table for many}
\end{figure}

\begin{figure}[h]
\resizebox{.25\textwidth}{!}{\includegraphics{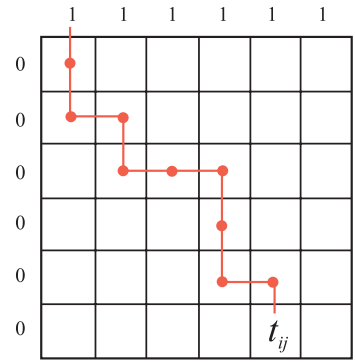}}
\caption{To determine the contribution of the element $t_{0k}$ into the element $t_{ij}$ one
has to find all possible paths connecting the two elements as long as only steps down or to the right are allowed.}
\label{Table for walks}
\end{figure}

Let us find the value of an element $t_{ij}$ for arbitrary $i$
and $j$. According to the recurrence relation
(\ref{recurrent}), any element is determined by its neighbors
that are immediately above and immediately to the left. An
element $t_{ij}$ therefore ultimately ``acquires'' its value
from those and only those initial elements $t_{0k}$ that have
$k\leq j$. Each such element $t_{0k}$ contributes into $t_{ij}$
the amount
\begin{equation}
\label{weight of every path}
\frac{t_0}{2^{i+j-k}} \times \frac{(i+j-k-1)!}{(i-1)! (j-k)!}.
\end{equation}
In the first factor of this expression the exponent $i+j-k$ is
equal to the number of steps it takes to reach the position
$ij$ starting from the position $0k$, provided that (i) only
steps down or to the right are allowed, and (ii) each step
connects two adjacent elements; see Fig.~\ref{Table for walks}.
Any step decreases the weight of the element $t_{0k}$ by a factor of
$1/2$. The second factor in Eq.~(\ref{weight of every path}) is
the number of distinct paths connecting the elements
$t_{0k}$ and $t_{ij}$, given by the standard combinatorics formula
for the total number of possible combinations of $i-1$ vertical
steps and $j-k$ horizontal steps.  Note that the number of
different vertical steps is only $i-1$ rather than $i$. This is
because the very first step from the element $t_{0k}$ can
be taken only in the downward direction.

\begin{figure}[h]
\resizebox{.40\textwidth}{!}{\includegraphics{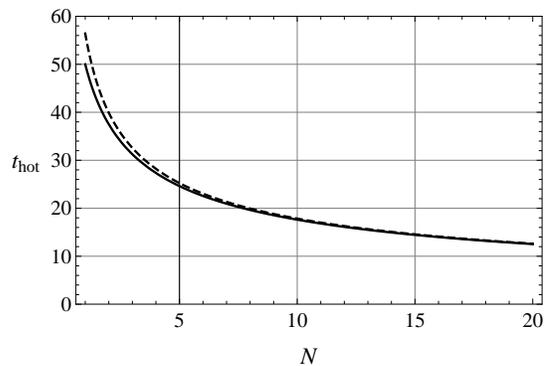} }
\caption{The solid line shows the exact $t_{\rm
hot}(N)$ given by Eq.~(\ref{final for N 2}). The dashed line demonstrates an excellent accuracy of the approximation
(\ref{large n limit}) for $N\gtrsim 5$.} \label{Plots}
\end{figure}

Summing the contributions (\ref{weight of every path}) for all
possible $0 < k\le j$, we obtain
\begin{equation}
\label{tij}
t_{ij} = \frac{t_0}{2^{i+j} (i-1)!} \sum_{k=1}^{j}2^k \frac{(i+j-k-1)!}{ (j-k)!}.
\end{equation}
The final temperatures can now be found from Eqs.~(\ref{final
temperature}).  To avoid interrupting the narrative, we move this
technical step to the appendix and present here only the
result,
\begin{equation}
\label{final for N 2} t_{\rm hot}(N)= t_0\frac{(2N)!}{4^N (N!)^2}.
\end{equation}
For a large number of splittings, $N \gg 1$, this expression can
be further simplified with the help of Stirling's approximation
for the factorials,
\begin{equation}
N!= \sqrt{2\pi N} \,(N/e)^N,
\end{equation}
leading to the approximation (\ref{large n limit}). The curves
in Fig.~\ref{Plots} show the accuracy of this approximation:
already for $N=5$ the expression (\ref{large n limit})
overestimates the exact result (\ref{final for N 2}) by only
about $2\%$.

We emphasize that while the final temperature of the hot
water tends to zero asymptotically for $N\to \infty$, it
approaches zero rather slowly. For example, it takes splitting
into $N=128$ parts to cool the water to below $5\,\degree \C$.
At least in principle, however, we can reverse the hot and cold temperatures
to any desired
degree by choosing a sufficiently large~$N$. (Of course, in
a realistic system thermal losses to the  environment would
limit the efficiency of the heat exchange.)

\section{The continuous limit, $N\to \infty$}

The complete exchange of temperatures in the
limit $N\to \infty$ can also be derived from a
continuum approach.  As the size of the fraction
$1/N$ tends to zero, the matrix element $t_{ij}$ can be viewed
as a progressively smoother function of $x=i/N$ and $y=j/N$,
which vary from $0$ to $1$, so that instead of a discrete
matrix we have a function of two variables,
\begin{equation}
t_{ij} = t_{xN,yN} \equiv t_N (x,y),
\end{equation}
which becomes continuous in the limit $N\to \infty$. In this
limit the final temperature can be written as an integral:
\begin{equation}
t_{\rm hot} =\frac{1}{N} \sum_{j=0}^N t_N(1,j/N) \to \int\limits_0^1 dy\, t_\infty (1,y).
\end{equation}
Similarly, the final temperature of the cold water is $t_{\rm
cold} =\int_0^1\, dx \,t_\infty (x,1).$

The function $t_\infty (x,y)$ can now be found from a
differential equation derived from the continuum limit of the
recursion relation (\ref{recurrent}). In the new notations for
the function $t_N (x,y)$ the recursion relations are
\begin{equation}
\label{recursion in continuous}
t_N (x,y)=\frac{1}{2} \Bigl(t_N (x-{1}/{N},y)+t_N (x,y-{1}/{N}) \Bigr).
\end{equation}
In the large-$N$ limit  the temperature function can be
expanded to the linear order in $1/N$:
\begin{equation}
\label{Taylor expansion}
t_N (x-1/N,y)=t_N (x,y) -\frac{1}{N} \frac{\partial t_N (x,y)}{\partial x} +O(1/N^2),
\end{equation}
and similarly for the second term in Eq.~(\ref{recursion in
continuous}). Disregarding the terms of the second and higher
orders in $1/N$, which vanish in the leading
approximation, we derive from Eq.~(\ref{recursion in
continuous}) the partial differential equation
\begin{equation}
\label{partial differential equation}
\frac{\partial t_\infty (x,y)}{\partial x} +\frac{\partial t_\infty (x,y)}{\partial y}=0.
\end{equation}
The expressions (\ref{zero row and column}) serve as the
boundary conditions for this equation; in the continuous
notations they become
\begin{equation}
t_\infty (0,y)=t_0,\hspace{0.5cm} t_\infty (x,0) =0.
\end{equation}

In order to solve Eq.~(\ref{partial differential equation}), we
first notice that any function of the \textit{difference} of
the two arguments satisfies it identically:
\begin{equation}
t_\infty (x,y)=f (x-y).
\end{equation}
Second, from the basic theory of differential equations it is
known that a general solution of a linear partial differential
equation contains one and only one arbitrary function
(similarly to how an ordinary linear differential equation
contains one arbitrary constant).

It is now clear that both the equation and its boundary
conditions will be satisfied by a solution proportional to
the Heaviside step function,
\begin{equation}
\label{partial differential equation solution}
t_\infty (x,y)=t_0\Theta(x-y),\hspace{0.5cm} \Theta(z)= \left\{\begin{array}{ll}1, & z>0,\\ 0,& z<0. \end{array}  \right.
\end{equation}
Note that this solution  is not well defined at coincident
arguments. In this somewhat special case we need to impose
manually that $t_\infty (1/2,1/2) =t_0/2$, to reflect the
corresponding property of the diagonal matrix elements for any
finite~$N$. This behavior is reminiscent of the Fermi-Dirac
distribution, which is step-like at zero temperature but whose
value is $1/2$ right at the chemical potential for any nonzero
temperature.\cite{LL}

From the step-like character of the solution (\ref{partial
differential equation solution}) the complete temperature
reversal becomes quite obvious: the ultimate temperature of the
hot water is $t_{\rm hot}(\infty) = t_0 \int_0^1 dy \,
\Theta(0-y)=0$.


\section{What about the Second Law of thermodynamics?}

The complete reversal of temperatures in the $N\to \infty $
limit is a surprising result that seems paradoxical at first
glance. Indeed, the return of the closed system to the initial
state---essentially the outcome in our case---is a signature
of a reversible process. But thermal equilibration between
objects with temperatures that differ by a non-infinitesimal
amount is expected to be irreversible.  Clearly that is not
what happens here, even though the difference in temperatures is
large.

The paradox  is resolved if one notices that the conventional
reasoning applies only to \textit{large} bodies. When the objects
themselves are infinitesimal, reversibility turns out to be
possible even for nonzero differences in temperatures.

To illustrate this, let us look at what happens to the entropy
of the system. We denote by $C_0$ the heat capacity of the whole
amount of each water. When the temperature of the water (cold
or hot) changes from some initial $T_i$ to a final $T_f$, its
entropy changes by \cite{LL}
\begin{equation}
\Delta S = C_0 \int_{T_i}^{T_f} \frac{dT}{T} =\frac{C_0}{N}\ln{\frac{T_f}{T_i}}.
\end{equation}
We emphasize that $T$ denotes the thermodynamic temperature,
$T= T_0 +t$, with $T_0=273$~K in our case.

The initial temperatures of the hot and cold waters are $T_0+t_0$
and $T_0$, respectively. At the end of the heat exchange, the
temperature of the ``hot'' water will go down to $T_0+t_{\rm
hot}$ while the ``cold'' water's will rise to $T_0+t_0 -t_{\rm
hot}$. The total increase of the system's entropy will thus be
\begin{equation}
\label{total entropy change}
\Delta S_{\rm tot} = C_0 \left(\ln \frac{T_0+t_{\rm hot}}{T_0}
+\ln \frac{T_0+t_0 -t_{\rm hot}}{T_0+t_0}\right).
\end{equation}
Because for large $N$ the value $t_{\rm hot}(N)\ll T_0$, one
can expand the arguments of the logarithms using $\ln(1+x)
\approx x$, and then simplify the result with the help of
Eq.~(\ref{large n limit}):
\begin{align}
\label{total entropy change 1}
\Delta S_{\rm tot}(N)
\approx& C_0 \left(\frac{t_{\rm hot}(N)}{T_0}-\frac{t_{\rm hot}(N)}{T_0+t_0}\right)\nonumber\\
= &\frac{C_0 t_0}{T_0(T_0+t_0)}t_{\rm hot}(N) \approx \frac{C_0}{
\sqrt{\pi N}} \frac{ t_0^2}{T_0(T_0+t_0)}.
\end{align}

\begin{figure}[h]
\resizebox{.40\textwidth}{!}{\includegraphics{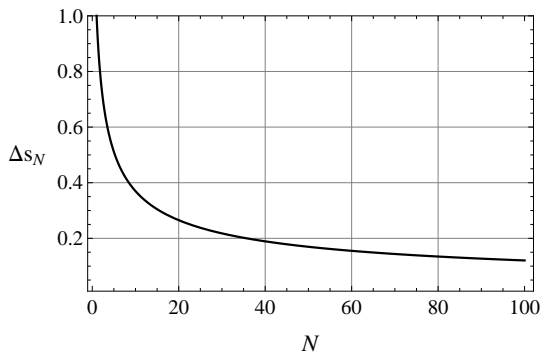} }
\centering
\caption{The total entropy change for $N$ splittings, relative
to the entropy produced when the waters are brought into contact without splittling,
$\Delta s_N=\Delta S_{\rm tot}(N)/\Delta S_{\rm tot}(1)$, as determined by Eq.~(\ref{total entropy change}).
} \label{PlotEntropy}
\end{figure}

Figure~\ref{PlotEntropy} illustrates the how the net increase in
entropy, Eq.~(\ref{total entropy change}), depends on $N$.  The
total change in entropy  vanishes in inverse proportion to the
square root of the  number of parts. Note that Eq.~(\ref{total
entropy change 1})
 proves reversibility in the limit of large $N$ in a strict mathematical sense\cite{Tho}:
 for any pre-assigned $\Delta \tilde S$, no matter how small, there is a number
 $\tilde N$ such that for $N>\tilde N$ the entropy increase will be less than the pre-assigned value:
$ \Delta S(N) < \Delta \tilde S$.

It is instructive to interpret Eq.~(\ref{total
entropy change 1}) in qualitative terms. When the first element
of cold water equilibrates with the first element of hot water,
the temperature difference is large and the entropy increases
by the maximum amount, $\sim C_0/N$. The resulting warming of
cold water and cooling of hot water ensures that, as further
elements are brought into contact, the temperature differences
decrease. This makes subsequent contacts less ``efficient''
producers of entropy. As formula (\ref{total entropy change})
suggests, by the order of magnitude, roughly $\sim \sqrt{N}$
contacts generate the maximum entropy, $\sim C_0/N$, meaning
that the total entropy increase is of the order $\sim
C_0/\sqrt{N}$, and vanishes in the limit of large $N$.

\section{Conclusions}

A complete  temperature reversal occurs when two bodies are
split into infinitesimally small parts and brought in thermal
contact sequentially. This example illustrates a type of reversible process that is less familiar than the
well-known scenario.  In that scenario one allows the bodies to have arbitrary  size but imposes
the condition that their temperatures differ by an
infinitesimal amount.  In the current illustration, to the
contrary, the temperature differences can be arbitrary,
but the bodies must be split into infinitesimal parts.

Needless to say, the analysis of this paper is subject to
the usual thermodynamic restrictions. First, we have assumed that it
is possible to neglect thermal losses to other bodies and to
the environment (via thermal radiation, for example). Second,
by ``infinitesimal'' parts we mean amounts of the
material that are still large enough to contain many microscopic particles; the
parts are thus infinitesimal only in the usual thermodynamic
sense.

On the other hand, the assumption of the
temperature-independent specific heat of water, while
convenient, is not crucial. Even when this
requirement is relaxed, the complete temperature reversal
still occurs.  Instead of the recurrence condition
(\ref{recurrent}), one has to satisfy the condition of
energy conservation. More exactly, if the heat exchange occurs
at a constant pressure $P$, one has to require the conservation
of the enthalpy $H(T,P)$ before and after
each thermal contact:
\begin{equation}
\label{Gibbs}
2H(T_{ij},P) = H(T_{i-1\, j},P) + H(T_{i\, j-1},P).
\end{equation}
Repeating now the arguments  for the limit of $N\to
\infty$ developed at the end of the previous section, we obtain the
continuous version of Eq.~(\ref{Gibbs}):
\begin{equation}
\label{partial differential equation Gibbs}
\frac{\partial H(T,P)}{\partial T} \left( \frac{\partial T_\infty (x,y)}{\partial x} +\frac{\partial T_\infty (x,y)}{\partial y} \right)=0.
\end{equation}
The derivative $\partial H /\partial T$ is the entropy, which is never zero for any nonzero~$T$. As a result,
the temperature function $T(x,y)$ satisfies exactly the same
equation as in the case of constant specific heat.
Correspondingly, this equation has the same solution (\ref{partial
differential equation solution}) leading to the same conclusion
of complete temperature reversal.

From a practical standpoint, the principle discussed in this
paper can be realized when heat is transferred between two
fluids through a dividing wall that prevents them from mixing---a
design known as a \textit{recuperator} or \textit{direct-transfer heat
exchanger}.\cite{exchanger} One obvious factor limiting the
efficiency of such an exchanger is the heat flow between different
parts of each liquid, which results in ``premature''
thermalization, excluded in our idealized scheme.

\begin{acknowledgments}

Discussions with Stephan LeBohec and Mikhail Raikh are
gratefully acknowledged.

\end{acknowledgments}

\begin{appendix}*

\section{Derivation of Eq.~(\ref{final for N 2})}

To prove our final result (\ref{final for N 2}) for an arbitrary
number of parts~$N$, we rewrite
Eq.~(\ref{tij}) in a slightly simpler form by replacing the
summation variable $k$ with the new index $l=j-k$:
\begin{equation}
\label{arbitrary tij}
t_{ij} = \frac{t_0}{2^i (i-1)!} \sum_{l=0}^{j-1} \frac{(i+l-1)!}{ 2^l l!}.
\end{equation}
A helpful identity follows from this formula and the second of
the properties (\ref{properties of t}). Substituting $i=j= N+1$
into Eq.~(\ref{arbitrary tij}), we obtain that  for any $N$,
\begin{equation}
\label{magic identity}
\sum_{l=0}^{N}\frac{(N+l)!}{ 2^l\, l!} =2^NN!.
\end{equation}

\begin{figure}[h]
\resizebox{.30\textwidth}{!}{\includegraphics{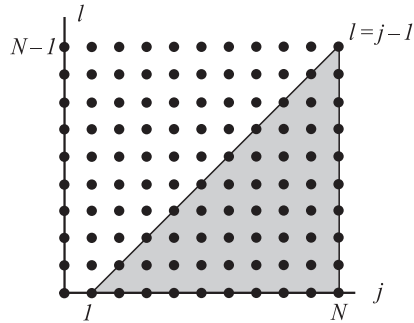}}
\caption{The summation in Eq.~(\ref{final for N}) is performed over the points belonging to the
shaded triangle. It can be done in two equivalent ways, with the summations over $l$ or $j$ performed first, as prescribed by
Eq.~(\ref{change of summations}).}
\label{Grid}
\end{figure}

Having determined all elements of the matrix $t_{ij}$, we can
now  calculate the final temperature $t_{\rm hot}(N)$. As
explained in the main text, it is given by the average of the
first $N$ elements of  $N$th row (see Eq.~(\ref{final
temperature})),
\begin{equation}
\label{final for N}
t_{\rm
hot}(N)= \frac{t_0}{2^N N!}
\sum_{j=1}^N\sum_{l=0}^{j-1} \frac{(N+l-1)!}{ 2^l l!}.
\end{equation}
Because the expression inside the summations depends on $l$ but
not on $j$, it is convenient to perform the summation
over $j$ first. To do so, we replace
\begin{equation}
\label{change of summations}
\sum_{j=1}^N\sum_{l=0}^{j-1} \to \sum_{l=0}^{N-1}\sum_{j=l+1}^N.
\end{equation}
The meaning of the change of the order of the summations is
illustrated in Fig.~\ref{Grid}. It is quite analogous to the
reversal of the order of multiple integrations in standard
calculus.

The $j$ summation now simply yields the total number of terms
in the sum, $N-l$:
\begin{equation}
\label{final for N 1}
t_{\rm
hot}(N)= \frac{t_0}{2^N N!}
\sum_{l=0}^{N-1} \frac{(N-l)(N+l-1)!}{ 2^l l!}.
\end{equation}
Now we must evaluate the remaining sum over~$l$, which we abbreviate
as $s_N$.
The first step is to separate the two terms in the
difference $N-l$:
\begin{align}
\label{appendix 1}
s_N =& \sum_{l=0}^{N-1}\frac{(N-l)(N+l-1)!}{2^l l!} \nonumber\\
=& N\sum_{l=0}^{N-1}\frac{(N+l-1)!}{2^l l!} - \sum_{l=1}^{N-1}\frac{(N+l-1)!}{2^l (l-1)!}.
\end{align}
Note that in
the last term the summation begins at $l=1$,
because of the (canceled) factor of $l$ in the numerator.
The first term on the right-hand side of Eq.~(\ref{appendix 1})
is of the form of the sum in the relation (\ref{magic
identity}) and is equal to $2^{N-1} N!$, while the  last term
in Eq.~(\ref{appendix 1}) can be brought into a similar form by
the substitution $l= k+1$:
 \begin{eqnarray}
\label{appendix 2}
s_N = 2^{N-1} N! -\frac{1}{2}\sum_{k=0}^{N-2}\frac{(N+k)!}{2^{k} k!}.
\end{eqnarray}
The remaining sum has almost the same form as the sum in  the relation (\ref{magic identity}), except it lacks the two largest terms. Those terms can be subtracted explicitly:
\begin{equation}
s_N = 2^{N-1} N! -\frac{1}{2}\left(2^{N}N!-\frac{(2N)!}{2^N N!} -  \frac{(2N-1)!}{2^{N-1} (N-1)!} \right).
\end{equation}
The first term inside the parenthesis cancels with the term
outside, while the remaining two terms are easily verified to be
equal to each other. Therefore, we obtain
\begin{equation}
s_N = \frac{(2N)!}{2^N N!}.
\end{equation}
Replacing the sum in Eq.~(\ref{final for N 1}) with this expression, we
obtain our main result (\ref{final for N 2}).

In principle, this should complete our solution. The only
remaining concern is the validity of relation (\ref{magic
identity}), which we obtained in an essentially ``empirical''
way, by observing that all diagonal components of the
temperature matrix are equal to $1/2$.  We now prove this relation rigorously.\cite{Raikh}

Consider the sum
\begin{equation}
\label{Raikh1}
Z_N =\sum_{l=0}^{N}\frac{(N+l)!}{2^l l!},
\end{equation}
and let us express the value $Z_{N+1}$ via $Z_N$:
\begin{eqnarray}
Z_{N+1} &=& \sum_{l=0}^{N+1}\frac{(N+l+1)!}{2^l l!}=\sum_{l=0}^{N+1}\frac{(N+l+1)(N+l)!}{2^l l!}\nonumber\\
&=& (N+1) \sum_{l=0}^{N+1}\frac{(N+l)!}{2^l l!}+\sum_{l=1}^{N+1}\frac{(N+l)!}{2^l (l-1)!}.
\end{eqnarray}
In the first sum we note that all the terms with the exception
of the last one with $l=N+1$ combine into $Z_N$. In the second
sum we replace $l=k+1$:
\begin{align}
Z_{N+1} =(N+1)\left(Z_N +\frac{(2N+1)!}{2^{N+1} (N+1)!} \right)\nonumber\\  + \frac{1}{2} \sum_{k=0}^{N}\frac{(N+k+1)!}{2^k k!}.
\end{align}
The last sum almost adds up to $Z_{N+1}$, but this time it
lacks the $k=N+1$ term, which can then be explicitly subtracted
from $Z_{N+1}$:
\begin{align}
Z_{N+1} =&(N+1)\left(Z_N +\frac{(2N+1)!}{2^{N+1} (N+1)!} \right) \nonumber\\&+ \frac{1}{2} \left(Z_{N+1}
-\frac{(2N+2)!}{2^N (N+1)!}\right).
\end{align}
The two terms with the factorials exactly cancel each other.
The remaining terms yield the identity
\begin{equation}
Z_{N+1} = 2(N+1)Z_N.
\end{equation}
Together with the initial condition $Z_0 =1$, this recurrence
relation gives $Z_N= 2^NN!$, thus proving Eq.~(\ref{magic
identity}).

\end{appendix}

\end{document}